\documentclass[letterpaper]{jpconf}
\usepackage{graphicx}

\begin{document}

\title{Observations of high energy neutrinos with water/ice neutrino 
telescopes.
}

  \author{
  Albrecht Karle    
  }
  \address{Department of Physics, University of Wisconsin, Madison, 
  U.S.A.}
  
  \ead{karle@icecube.wisc.edu}

\begin{abstract}
The search for high energy neutrinos of astrophysical origin is 
being conducted today with two water/ice Cherenkov experiments. 
New instruments of higher performance are now in construction 
and more are in the R\&D phase. 
No sources have been found to date.  Upper limits on neutrino fluxes are approaching 
model predictions.  Results are reported on the search for point sources, 
diffuse fluxes, gamma ray bursts,  dark matter and other sources. 
\end{abstract}

\section{Introduction}
\label{sec:intro}

High energy neutrino astronomy will offer a new view of the 
Universe.  The observation of the highest energy cosmic rays and 
the detection of galactic and extragalactic sources of 
gamma rays are proof that very energetic sources
of non-thermal radiation exist. 
The detection of sources of high energy neutrinos
will provide insights into the origin of the highest energy radiation. 
Neutrinos have been detected from the Sun and from Supernova 1987a; 
both observations representing fundamental breakthroughs in  
astrophysics and physics. 
Neutrino telescopes now in operation and construction aim at higher
energies.  This report will focus on the energy range 
aimed for by under water/ice neutrino telescopes: $\approx 10^{11}$\,eV to 
$\approx 10^{18}$eV. 
At the same time an increasing effort is underway to detect 
neutrinos at the highest energies to beyond $10^{20}$\,eV.
Markov and Zheleznikh \cite{mar} suggested the 
detection of neutrinos in water  via 
the process $ \nu_l(\bar{\nu_l})+N \rightarrow l^{\pm} + X   $
of upward or horizontal neutrinos interacting with a nucleon $N$ of the 
matter surrounding the detector.  A significant fraction of the 
neutrino energy will be carried away by the produced lepton. 
For the important case of the muon the angle between the 
parent neutrino and the muon is less than $1^{\circ}$ at 1 TeV energy and it becomes 
very small at higher energies providing the basis for astronomy.

In the past decade high energy neutrino astronomy has evolved from 
exploratory experiments to the first generation of neutrino 
telescopes with substantial neutrino detection rates.
The first attempt was made
by the DUMAND project \cite{dumand} in the deep Pacific Ocean near Hawaii. 
This pioneering effort
opened the road to the realization of the first working 
neutrino telescopes, AMANDA-B10 (1997, \cite{b4}),  Baikal (NT-200 1998, 
\cite{baikal})
and AMANDA-II  (2000). 
Two telescopes are in construction, IceCube \cite{icecube-icrc}, at the South Pole, 
and ANTARES \cite{antares}, in the Mediterranean Sea, and two projects 
are in the planning and 
prototyping phase in the Mediterranean Sea: NESTOR \cite{nestor} and NEMO.  
The sensitivity of some of the current experiments, Baikal and 
AMANDA-II, is already below some model predictions. This report
will focus on the current status of observations.

\section{Detection principle and energy ranges}

Water Cherenkov detectors detect the Cherenkov photons 
that are emitted by relativistic particles.   
Strings of optical sensors which are deployed in a transparent medium
detect the Cherenkov photons of muon tracks or cascades. 
Measuring the photon arrival times with a resolution of a few nanoseconds 
allows an accurate reconstruction of the muon track. 

\begin{figure}
  \begin{center}
    \begin{minipage}[c]{0.6\linewidth}
 \includegraphics[width=0.98\textwidth]{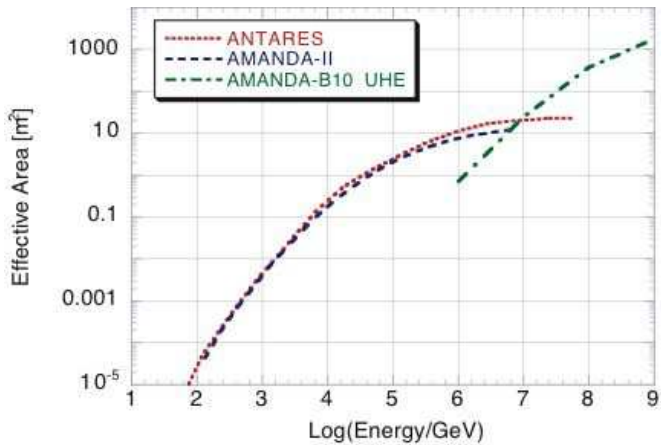}
  \end{minipage}\hfill
  \begin{minipage}[t]{0.36\linewidth}
 \vskip -3cm
	\caption{Neutrino effective areas of 
  AMANDA-II obtained
with the cuts of the point-like source analysis (2000-3, 
\cite{icecube-icrc})
and for the AMANDA B-10 analysis on UHE energy events
(\cite{amanda-b10-uhe}), and for ANTARES \cite{montaruli}.
The $\nu$-effective 
 area of neutrino telescopes increases strongly with energy. 
  The neutrino effective area is defined as the the equivalent area 
for which the neutrino detection probability would be 100\%. 
 \label{fig:nu-area}}
    \end{minipage}
  \end{center}
\end{figure}

Neutrino telescopes cover a wide energy range for 
two reasons. The first reason is that the neutrino-nucleon cross section
increases substantially with energy.  The second is that the muon range 
and energy loss increase substantially with energy.
We single out muons because they allow the best angular 
resolution.  The cascade channel (e's, $\tau$'s 
and neutral currents) however allow in many areas a 
comparable sensitivity. 
Figure \ref{fig:nu-area} shows the effective detection area for muon neutrinos 
in AMANDA-II and ANTARES \cite{montaruli} for point source selection, 
for the Ultra High Energy (UHE) analysis of AMANDA B10.  

\section{Detectors}

\subsection{AMANDA}

\begin{figure}
 \includegraphics[width=0.95\textwidth]{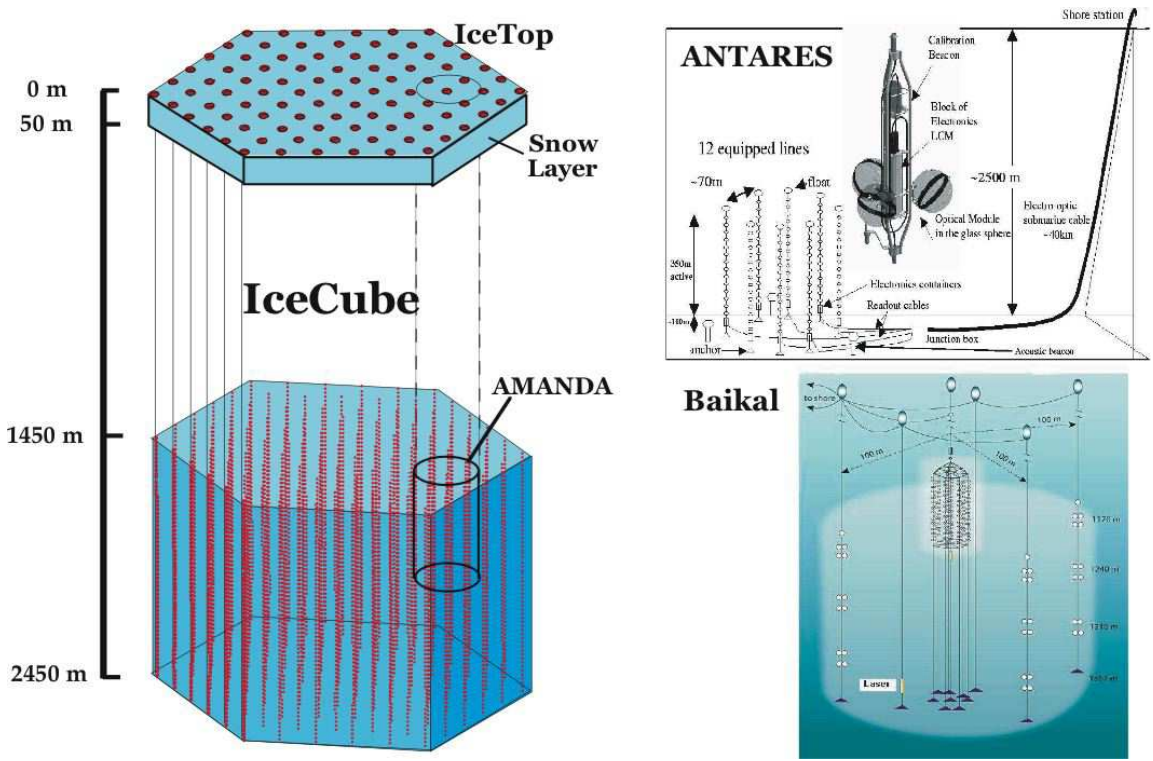}\hspace*{\fill}
\caption{Schematic view of currently running detectors (AMANDA and 
Baikal) and of detectors currently 
in construction:  ANTARES and IceCube.}
\end{figure}

AMANDA is a neutrino telescope built and operated at the
Geographic South Pole. The final detector configuration, called AMANDA-II, 
consists of 677 optical modules (OM)
arranged on 19 vertical strings deployed at depths between 
1300-2400\,m. 
Most of the OMs are deployed at a depth of 1500-1950\,m and arranged 
within a diameter of 200\,m. 
All analog pulses of the 20\,cm photomultipliers are transmitted 
to the surface on individual 
cables, with most of the channels using fiber optic cables. 
The analog transmission on fibers allows low dispersion
(PMT pulse FWHM about 20 nsec), high time resolution and a high dynamic range
while keeping the electronics in the ice at a minimum.

The IceCube neutrino observatory at the South Pole will consist of 
4800 optical sensors, installed on 80 strings between the depths of
1450\,m to 2450\,m in the Antarctic Ice, and 320 sensors deployed in 
160 ice Cherenkov tanks (referred to as IceTop)
on the ice surface directly above the strings. Each sensor consists 
of a 25\,cm photomultiplier
tube, connected to a waveform-recording data acquisition circuit capable
of resolving pulses at nsec time resolution and high dynamic range. 
In January 2005 76 such sensors were installed as a first
part of the IceCube array. 
The collaboration reported first results which show that the detector 
system  works as expected \cite{icecube-icrc}.  
The collaboration hopes to do first science with an array of 
10 or more strings in 2006 and complete construction in 2010/11.

\subsection{Baikal}

The Baikal neutrino telescope is located 30\,km off the 
shore of Lake Baikal, Siberia, at a depth of 1.1 km.
The configuration NT200 \cite{baikal} 
was commissioned in April, 1998.
It consists of 192 optical modules (OMs), mounted 
on 8 vertical strings, which form a structure 
of 40\,m diameter and 72\,m height.
Each OM contains a 37\,cm diameter photomultiplier.
For installation and maintenance, the Baikal collaboration takes 
advantage of the cold winters in Siberia, which 
cause the Lake to freeze over for several months allowing easy access 
to the deployment site. 
The Baikal Collaboration added three outer strings in 
2005 to increase the sensitivity at high energies for cascades.

\subsection{ANTARES}

Antares is a neutrino telescope under construction  
in the Mediteranean Sea at a depth of 2500\,m near 
Toulon, France.  
The project aims to complete the full array, consisting of 12 
lines with 75 photomultipliers each, by 2007. 
Simulations indicate a very good angular resolution 
of $ 0.3^{\circ} $ for 1 TeV $\mu's$ ($ 0.6^{\circ} $ degrees with respect to the 
neutrino direction).  
The neutrino effective area is expected to be similar to 
the AMANDA array as shown in fig. \ref{fig:nu-area}. 

\subsection{Other projects and initiatives}

NESTOR is a neutrino telescope with a proposed location near 
Pylos,  on the Greek Ionian Sea coast. 
A calibration and
engineering run of a test detector was carried out in 2003 
\cite{nestor-operation}. The detector was operated for more
than one month and data was continuously transmitted to shore.
With the data collected, the collaboration was able to 
reconstruct muon events at a rate consistent with predictions and 
produce a muon zenith angle distribution \cite{nestor-measure}.

The initial goal of the NEMO project is research and development to 
lay the technological foundation for a future km scale detector in 
the Mediterranean. 
The collaboration has identified a possible site for a $1\,km^{3}¥$-scale 
instrument at a depth of 3500\,m, at a distance of about 
80\,km from the Sicilian coast. 
The NEMO architecture for a $1\,km^{3}¥$ detector is based on 5832 optical 
sensors.
Finally, it should be mentioned that the groups involved in the 
Meditarranean projects have formed a European consortium called KM3Net 
to further the development of a kilometer scale 
neutrino telescope in the Mediterranean Sea.

\section{Atmospheric muons and neutrinos}

Cosmic ray muons form a background for neutrino telescopes.  
The background is rejected in two ways which illustrate two 
fundamental modes of operation:
\begin{enumerate}
\item At low energies - below a few PeV:  Reject downgoing muons by 
zenith angle and use upward muons as reliable neutrino messengers. 
\item At very high energies:  Reject cosmic ray muon background by 
direction and energy.  At energies above $\approx 10^{15-16}$\,eV, 
(depending on the depth of the instrument) the down going 
background is negligible, especially at larger zenith angles.
This allows to accept more  downgoing signal events. 
\end{enumerate}
At the depth of AMANDA it is found that the down-going cosmic ray muon 
flux is about 5 orders of magnitude larger than the expected neutrino 
flux. 
At greater depth the muon flux is significantly smaller and vice versa. 
Neutrino telescopes can use the cosmic ray muon flux to
study the detector response, possible systematic effects and also to
perform physics measurements.  
The cosmic ray muon angular and
vertical intensities were measured in AMANDA-II and they agree well
with simulations and with other experimental measurements \cite{icecube-icrc}. 
AMANDA was able to use down-going muon events,
which are in coincidence with SPASE surface array, 
to make a measurement of the cosmic ray mass composition as a function of the
primary energy near the knee \cite{comp}.

In order to measure the atmospheric neutrino component, a 
rejection factor of more than $\sim 10^6$ is needed to eliminate the
cosmic ray muon background. 
The AMANDA collaboration records about 4 well reconstructed
neutrino events per day in AMANDA-II. 
The preliminary unfolded
$\nu_{\mu}$ sea-level energy spectrum was determined and is in
agreement with the expectations up to above $10^4$ GeV \cite{nunfold}.
The observed spectrum is shown in figure \ref{fig:diffuselimits}.

\section{Search for sources of astrophysical neutrinos} 

\subsection{Diffuse fluxes} 

The search for diffuse astrophysical neutrino fluxes is in many respects the 
most challenging.  
The reason is that atmospheric neutrino background rejection is almost entirely based on 
energy reconstruction, and misreconstructed downgoing cosmic ray muons can complicate 
the analysis.  
A diffuse flux can only be measured if the spectrum is significantly 
harder than the steeply falling atmospheric neutrino spectrum. 

Models of cosmic ray shock acceleration based on the Fermi mechanism 
naturally produce energy spectra with a power index $ \gamma \approx 2$. 
$E^{-2}$-type spectra
are often used as a reference flux. 
However, each specific model needs to be optimized and analyzed separately and 
the energy cut will move higher or lower depending on the 
hypothetical signal flux.  The limits discussed here are not 
differential limits, instead they refer to an assumed $E^{-2}$ spectrum 
extending over the whole energy range. 
The accelerated protons interact with each other or with ambient photons. Pion 
decay yields a $\nu-$flux with roughly
the same spectral index as the protons. At the production site, the 
$\nu-$flavors are distributed according
to the ratios $\nu_{e} : \nu_{\mu} :\nu_{\tau} = 1 : 2 : 0$. 
Due to neutrino oscillations, an 
approximate equipartition of flavors is predicted to be observed at 
the detector.
However, the assumption of a $\nu_{e} : \nu_{\mu} :\nu_{\tau} = 1 : 1: 
1$ ratio is indeed a simplification, which is not accurate at energies 
above $\approx 10^{14}¥\,eV $ \cite{waxman-ratios}.

\begin{figure}[htb]
\centering
 \includegraphics[width=0.7\textwidth]{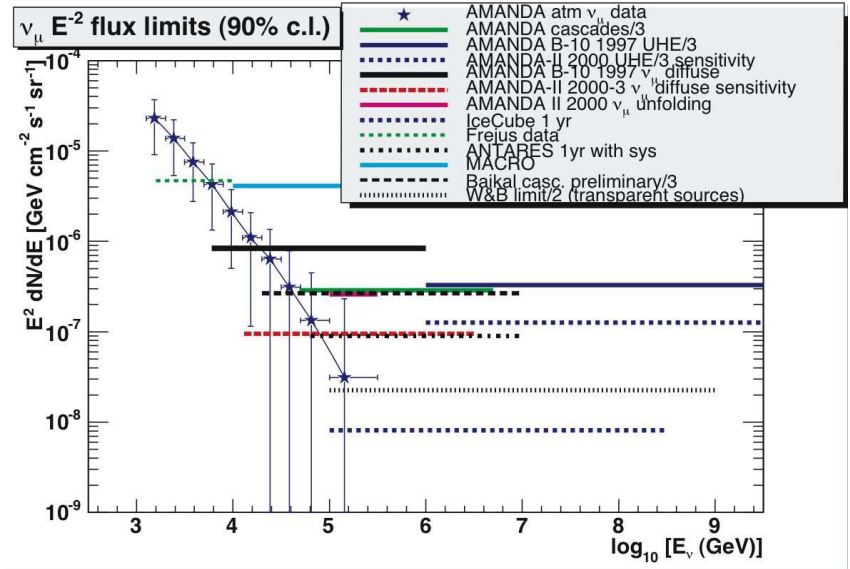}
\caption{Atmospheric neutrino flux, limits and sensitivities to 
astrophysical fluxes of an $E^{-2} $ energy spectrum.  
Limits by two underground detectors are shown: a) Frejus 
and b) MACRO \cite{macro}
The stars 
represent a measurement of the atmospheric 
neutrino flux by AMANDA-II. 
\cite{nunfold}.  Limits for 
$\nu_{\mu}+\nu_{\mu}$ fluxes are shown for 
a) AMANDA B10, 10 string array, 97 data 
\cite{diffb10}, b)
AMANDA-II unfolded spectrum 
\cite{nunfold}.
The sensitivity is shown for ANTARES 
\cite{antares-1year}, 
and for AMANDA-II, 2000-2003 data
\cite{icecube-icrc}.
Limits based on cascades are divided by the number of 
contributing flavors.  They are shown for AMANDA-II 
\cite{amanda2-cascadelimit} and for non-contained cascades
limits are shown by AMANDA-B10
\cite{amanda-b10-uhe} and by Baikal (preliminary)
\cite{baikalcascades}. 
}
\label{fig:diffuselimits}
\end{figure}

We categorize the results in three groups and give the 
current results for Baikal and AMANDA. 
The reported results are compiled in the figure \ref{fig:diffuselimits}. 
    AMANDA has reported results on all three 
    channels, Baikal on the third one.  
\begin{itemize}
    \item Through-going upward muon events:  \\
    For the through-going upward muons AMANDA reports results in two 
analyses.  In one analysis, an energy cut is applied and
observed events above the energy cut are compared to the 
signal and background prediction.  The optimization is done 
before the data is unblinded. 
The collaboration reported a sensitivity at the level of 
$\approx 10^{-7}¥$GeV/(cm$^{2}$¥s sr  \cite{icecube-icrc}
for a data set currently being analyzed.   
The other result is based on a neutrino energy spectrum analysis with a 
differential limit
at about 30 TeV based on an energy unfolding method \cite{nunfold}.
    \item Contained cascade events: \\
This analysis aims for the measurement of the diffuse flux
of $\nu_{\e}$, $\nu_{\tau}$ and due to neutral current interactions.
In order to determine a generic astrophysical flux limit 
the assumption is made that astrophysical 
neutrino fluxes appear in a  $\nu_{e}:\nu_{\mu}:\nu_{\tau}$=1:1:1 
ratio. While the directional resolution is relatively poor, 
a comparable flux limit is obtained due to the 
better energy resolution for cascades compared to $\nu_{\mu}$. 
    \item Not through-going muon and not contained cascade events:   \\
    The Baikal collaboration was able to develop an analysis 
    technique which allowed the effective detector volume 
    for very bright cascade events to increase substantially over the geometric 
    volume.  Even though non contained events are accepted with
    a vertex up 200\,m outside the instrument, the technique allows
    to reject downgoing background effectively by using a combination of energy and 
    directionality of the timing pattern.  The atmospheric 
    neutrino background is easily rejected based on energy. 
    An upper limit is reported based on 806 days  of livetime.  
    AMANDA developed a technique to detect non-contained events of 
    very high energy.  In this analysis  the 2 billion cosmic ray muon events 
    are rejected using the energy.  This UHE analysis is sensitive 
    to signal flux in the region from about 10\,PeV to 1000\,PeV.
\end{itemize} 



\subsection{Search for point sources} 

The detection of astrophysical point sources of neutrinos 
is perhaps the most prominent and also name bearing goal of neutrino astronomy. 
Numerous sources of TeV energy gamma rays have been detected in the 
past decade. Neutrino telescopes not only overlap at low energies 
with the energy 
range of current ground based gamma ray telescopes, but also extend 
in sensitivity to much higher energies. 

Results are available from the underground detectors MACRO and 
SuperKamiokande and also from Baikal and AMANDA.

\begin{figure}[htb]
 \centering
  \includegraphics[width=0.95\textwidth]{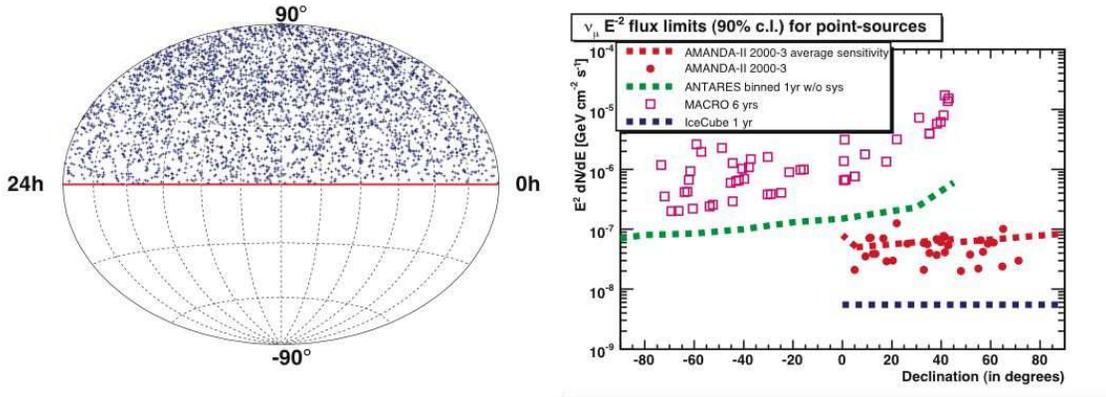}
\caption{Limits on E$^{-2}$-neutrino fluxes for point sources of 
 high energy neutrinos are shown for MACRO and AMANDA.  
 Also shown are average sensitivities for ANTARES and IceCube.}
 \label{fig:skyplot}
 \end{figure}
 
None of the current observations have indicated any 
significant excess above the background of atmospheric neutrinos. 
Limits are reported for individual sources
of special interest, which are also sources 
of gamma rays at TeV energies. 
Figure \ref{fig:skyplot} shows a number of individual limits 
for point sources from 4 years (807 days of livetime) 
of AMANDA in the northern hemisphere.  The limts are 
at the level of E$^{2}\cdot6\cdot 10^{-7}$¥GeV cm$^{-2}$sec$^{-2}¥$ \cite{icecube-icrc}. 
In a few cases, for example the active galaxy Mrk 501,
the limits on neutrino fluxes approach the observed 
gamma ray flux at TeV energy of these sources in their high states. 
Also shown are limits of the MACRO experiment from more than 10 years of 
data.
The average sensitivity for point sources with an $E^{2}$-spectrum 
is shown for the lifetime of MACRO, for 
4 years of AMANDA and for one year of IceCube.

AMANDA reported an upper limit from a search for neutrinos from the 
galactic plane in the Northern hemisphere \cite{icecube-icrc}.  
A larger km scale neutrino telescope will be needed to detect the 
a neutrino flux from the galactic plane.  The Northern 
hemisphere large instrument will be ideally positioned to search 
for the higher flux in near the 
galactic center. 

\subsection{Neutrinos from Gamma Ray Bursts, dark matter and other 
searches} 

Gamma-ray bursts (GRBs) are among the most energetic phenomena in the universe.
The observation of high energy neutrinos in gamma ray bursts 
would confirm hadronic acceleration in the fireball; 
possibly revealing an acceleration
mechanism for the highest energy cosmic rays.
AMANDA reported searches for neutrinos from a total 451 GRBs, most of which were 
triggered by the BATSE telescope  
 during 1997 to 2000. 
The short duration, typically 10 to 100 seconds, of the bursts makes 
background rejection easy.  
AMANDA observed zero neutrino events at an expected 
background of approximately 2 events for all bursts.  
Upper limits have been reported at the level 
of 3 $\cdot$ 10$^{-8}$ GeV¥cm$^{-2}$¥sr$^{-1}$¥s$^{-1}¥$ 
\cite{grb-rellen,icecube-icrc} for 139 bursts from 2000 to 2003. 
This limit is about an order of magnitude above the 
Waxman-Bahcall prediction \cite{waxmanbahcall-grb}.

\subsection{Dark matter and other searches} 

Minimally Supersymmetric extensions to the Standard Model predict the 
existence
of the neutralino, in the mass range GeV-TeV, which is a candidate for
the cold dark matter.  The particles will lose energy and become
gravitationally trapped in the centers of object like the earth, the
sun, or the center of the galaxy
where they annihilate to produce neutrinos.  These neutrinos may
be detected in a high-energy neutrino telescope.  
The energy range of the predicted neutrino fluxes is 
from about 10$^{10}$¥eV to a few times 10$^{12}$eV. 
AMANDA has performed a search for dark matter in the center of 
the Earth 
and the Sun 
\cite{darkmatterSunAmanda}.  No excesses of
events have been observed and upper limits were placed. 
Baikal reported an upper limit (see for example 
\cite{baikal-darkmatter}) at a comparable level on the neutrino 
flux from WIMPs annihilating in the Earth.  
SuperKamiokande holds the lowest limit by a factor of 2
\cite{superk-darkmatter} on the muon neutrino flux from the Sun at a level of 
2000\,km$^{-2}$¥yr$^{-1}$ for neutralino masses around 10$^{12}$eV
(based on a data set from many years).  Limits from underground 
detectors like Baksan \cite{baksan-darkmatter} and MACRO \cite{macro-darkmatter} 
are still comparable or lower, especially for lower neutralino masses 
around 10$^{11}$eV. 
ANTARES is expected to have a sensitivity of a factor 10 below the current 
limits;
IceCube is expected to have a final sensitivity of about 2 orders of magnitude below 
the current AMANDA limit. 
Nothern hemisphere telescopes 
will be in a good position to also search for dark matter from 
the center of the Galaxy. 

\subsection{Neutrinos of highest energies} 
New techniques are being explored to search for neutrinos 
in the energy range from $10^{18}$¥eV to beyond $10^{21}$¥eV. 
Some exploratory experiments are already underway.  A recent overview of sensitivities 
of techniques and instruments can be found in ref. \cite{saltzberg}. 
Most of the current efforts have focused on the use of radio 
signals from highest energy events in ice.  The application
of the radio technique in salt domes is
being investigated.  Acoustic techniques are being explored to 
measure the acoustic pulse of GZK events.  
First data are available from a prototype version of ANITA,
which has performed a first scan of the Antarctic 
ice sheet for neutrino radio pulses from a balloon \cite{anita}. 
RICE \cite{rice} uses radio sensors deployed in the deep ice 
near the South Pole.   
The techniques are still in rapid development and a substantial 
increase in sensitivity at highest energies 
can be expected in the next decade. 

\section*{Acknowledgments}
This work was supported by the NSF under OPP-0337726 grant.  The  
author thanks T. Montaruli, D. 
Boersma, G. Hill and S. Grullon for helpful comments.

\end{document}